\documentclass[runningheads]{llncs}
\usepackage[T1]{fontenc}
\usepackage{graphicx}
\usepackage{hyperref}       
\usepackage{url}            
\usepackage{booktabs}       
\usepackage{amsfonts}       
\usepackage{nicefrac}       
\usepackage{microtype}      
\usepackage{algorithmic}
\usepackage{algorithm}
\usepackage{amsmath}
\usepackage{graphicx}
\usepackage{caption}

\usepackage{pgfplots}

\usepackage{wrapfig}

\usepackage{mathtools}

\usepackage{caption,subcaption}

\begin{document}
\title{Improving Collaborative Filtering Recommendation via Graph Learning}
%
%
\author{Yongyu Wang}
\authorrunning{Y. Wang.}
%

\institute{JD Logistics\\
\email{wangyongyu1@jd.com}}

\maketitle              
\begin{abstract}
Recommendation systems aim to provide personalized predictions by identifying items that are most appealing to individual users. Among various recommendation approaches, k-nearest-neighbor (kNN)-based collaborative filtering (CF) remains one of the most widely used in practice. However, the kNN scheme often results in running the algorithm on a highly dense graph, which degrades computational efficiency. In addition, enforcing a uniform neighborhood size is not well suited to capturing the true underlying structure of the data. In this paper, we leverage recent advances in graph signal processing (GSP) to learn a sparse yet high-quality graph, improving the efficiency of collaborative filtering without sacrificing recommendation accuracy. Experiments on benchmark datasets demonstrate that our method can successfully perform CF-based recommendation using an extremely sparse graph while maintaining competitive performance.

\end{abstract}
\section{Introduction}
\label{sec:intro}

With the rapid growth of e-commerce, online shopping has become an integral part of daily life. The overwhelming volume of available products makes it infeasible for users to manually explore all options on a platform. As a result, algorithms that can accurately predict users' potential purchase interests and provide personalized recommendations are essential. Recommendation systems address this challenge by inferring users' preferences for items they have not yet interacted with, leveraging historical feedback and behavioral data. In this context, both predictive accuracy and computational efficiency are key objectives.

Collaborative filtering (CF) is among the most widely adopted approaches for recommendation. The central assumption is that users with similar historical behaviors tend to exhibit similar preferences: if two users have shown comparable tastes in the past, an item strongly favored by one user is likely to be favored by the other. In a typical user-based CF framework, to estimate a target user's preference for an unrated item, the algorithm first identifies the \(k\) most similar users according to historical rating patterns, and then predicts the target user's rating based on these neighbors' ratings of the item. From this perspective, CF can be viewed as a graph-based method that performs neighborhood aggregation (or label propagation) on a \(k\)-nearest-neighbor graph constructed from user--user similarities.

The performance of graph-based algorithms depends heavily on the quality of the underlying graph. The $k$-nearest-neighbor (kNN) graph, however, has several drawbacks that can directly limit the performance of collaborative filtering (CF). First, the optimal neighborhood size $k$ is problem-dependent and often difficult to determine. To adequately capture the underlying manifold of the data, $k$ typically needs to be sufficiently large; yet a large $k$ inevitably leads to a dense graph, which substantially increases the computational cost of graph-based CF. Moreover, in common multi-density scenarios, a fixed-size neighborhood is inherently inappropriate: a small $k$ may fail to cover meaningful local relations for many vertices, whereas a large $k$ may introduce redundant or even misleading edges for others. Therefore, an ideal graph for CF-based recommendation should both faithfully capture the underlying manifold structure and remain sufficiently sparse, so as to maintain recommendation quality while improving computational efficiency.

To this end, in this paper, we treat each item's \emph{user rating vector} as a graph signal and interpret collaborative filtering~(CF) as the propagation of graph signals over a user-network topology. From this perspective, we propose a graph signal processing~(GSP)--based method to learn sparse user networks from user--item interaction data. Our approach enables CF recommendation algorithms to operate efficiently on a much sparser graph without sacrificing recommendation accuracy.


\section{Preliminaries}
\subsection{Collaborative Filtering Recommendation Algorithms}

Let $U=\{u_1,\ldots,u_n\}$ denote the set of users and $I=\{i_1,\ldots,i_m\}$ the set of items. From historical feedback (e.g., explicit ratings, textual reviews, clicks, or purchase logs), we construct an $n\times m$ user--item interaction matrix $\mathbf{R}$, where the $j$th row represents user $u_j$ and the $\ell$th column represents item $i_\ell$. The entry $R_{j\ell}$ records the observed interaction between $u_j$ and $i_\ell$ (when available).

In user-based collaborative filtering (CF), the goal is to estimate the preference of a target (``active'') user $u_a$ for a target item. A typical procedure contains two stages:

\noindent\textbullet\ \emph{Neighborhood construction.}
Using the interaction vectors (rows of $\mathbf{R}$), the method evaluates how close $u_a$ is to each other user under a chosen similarity/distance measure (e.g., cosine similarity or Euclidean distance). The top-$k$ most similar users are then selected to form the neighborhood of $u_a$.

\noindent\textbullet\ \emph{Preference estimation.}
Let $\{nn_1,\ldots,nn_k\}$ be the selected neighbors of $u_a$, with similarity weights $\{s_1,\ldots,s_k\}$. For a target item $i_a$, let $r_i$ be the rating on $i_a$ provided by neighbor $nn_i$ (typically considering only neighbors who have rated $i_a$). The prediction $P_{u_a,i_a}$ can be computed via a similarity-weighted aggregation:
\begin{equation}\label{eqn:weightedsum}
P_{u_a,i_a} = \frac{\sum_{i=1}^{k} s_i\, r_i}{\sum_{i=1}^{k} s_i}.
\end{equation}
If $\sum_{i=1}^{k} s_i = 0$ or none of the neighbors has observed feedback on $i_a$, one commonly resorts to a back-off estimate such as a user mean, an item mean, or the global average.

\subsection{Graph Laplacian Matrices}
Consider a graph $G=(V,E,w)$, where $V$ is the vertex set of the graph, $E$ is the edge set of the graph, and $w$ is a weight function that assign positive weights to all edges. The Laplacian matrix of graph G is a symmetric diagonally dominant (SDD) matrix defined as follows:
\begin{equation}\label{formula_laplacian}
L_G(p,q)=\begin{cases}
-w(p,q) & \text{ if } (p,q)\in E \\
\sum\limits_{(p,t)\in E}w(p,t) & \text{ if } (p=q) \\
0 & \text{ if } otherwise.
\end{cases}
\end{equation}

According to spectral graph theory~\cite{chung1997spectral}, many important properties of a graph's topology are encoded in the spectrum of its Laplacian. 
Recent research~\cite{wang2022scalable,deng2022garnet} has further shown that manipulating graph edges via Laplacian-based spectral-domain analysis can improve the performance of downstream graph algorithms, such as spectral clustering and graph neural networks.

\section{Method}
In this section, we first propose the desirable properties that an ideal graph for CF-based recommendation should possess, and then present an algorithm to learn such a graph.

\subsection{Problem Formulation}

In CF, user preferences are inferred from observed user--item interactions, where the relational structure among entities can be encoded as a similarity graph.
In our implementation, we construct a user--user graph and instantiate the data matrix as the (observed) interaction matrix $X = R \in \mathbb{R}^{N \times M}$, where $N$ is the number of users, $M$ is the number of items, and $R_{u,i}$ denotes user $u$'s observed feedback (e.g., rating) on item $i$.
Under this instantiation, each column of $X$ represents one item's feedback collected from all users, and can be viewed as a graph signal on the graph. In this paper, we propose that the learned user--user graph should achieve the following characteristics:

\textbf{Smoothness of Graph Signals. }
We assume that users who are connected (i.e., deemed similar by the user--user graph) tend to exhibit similar preference patterns, so the per-item feedback observed across users should vary gradually along the graph rather than oscillate sharply between neighboring users. 
This ``neighbor-consistency'' assumption is exactly what makes graph-based CF work: when the feedback signal for an item is smooth on the user graph, a user's unknown preference can be reliably inferred from nearby users, because nearby nodes carry mutually informative signals.

\textbf{Sparsity of the learned Graph.} Graph sparsity is a critical consideration for CF-based recommendation. In our setting, the learned user--user graph is used directly in the CF inference process; therefore, a dense graph significantly amplifies the computational cost of neighborhood aggregation and graph-based signal propagation. A desirable graph-learning method should thus capture the essentialmanifold structure of the data using as few edges as possible. In this way, sparsity improves the scalability of CF without compromising the ability of the graph to support accurate preference prediction.

\subsection{Algorithm}

\begin{figure*}[!htbp]
\centering\includegraphics[scale=0.3]{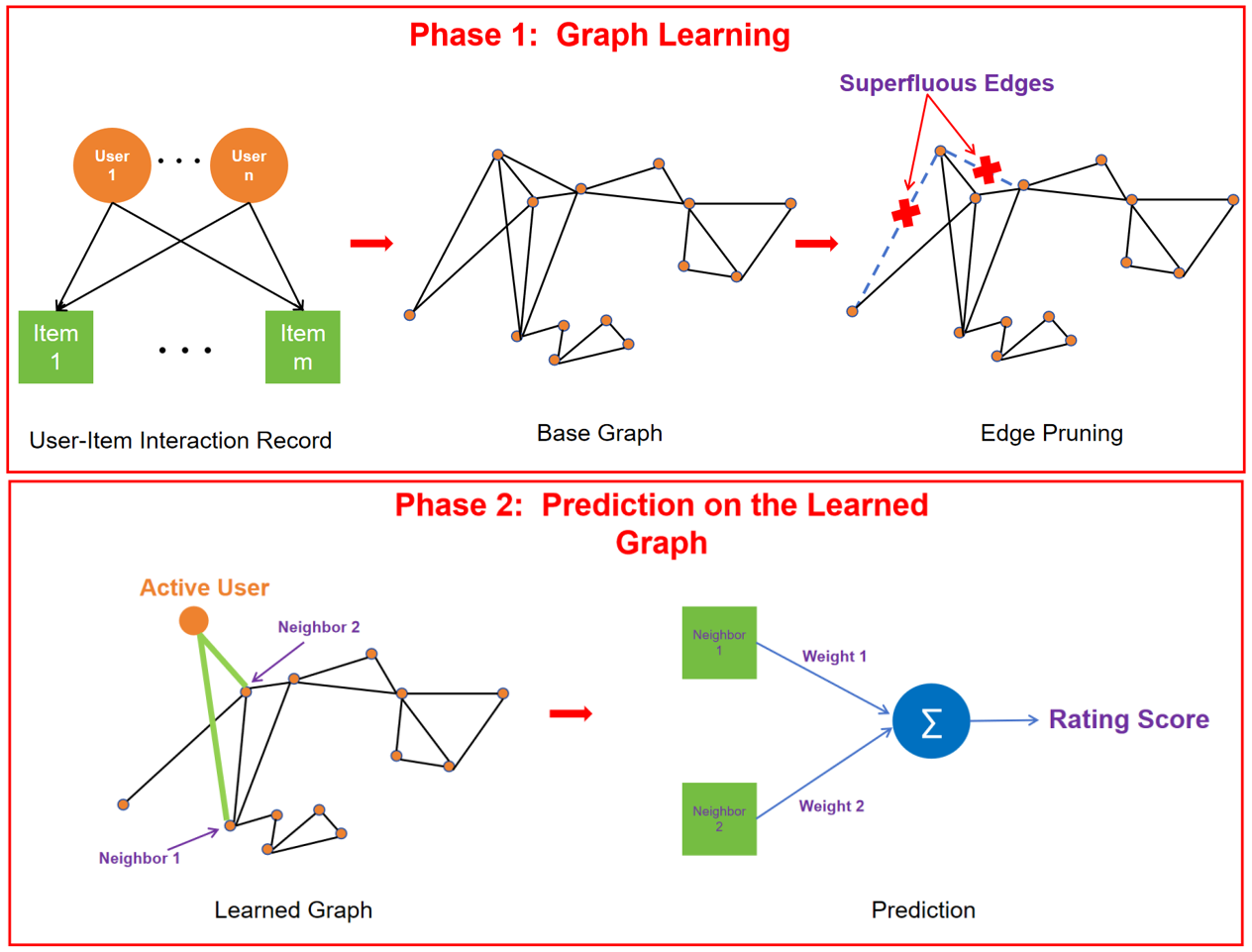}
\caption{An overview of the two major phases of the proposed method.\protect\label{fig:flowCF}}
\end{figure*}

Figure~\ref{fig:flowCF} illustrates the proposed method for efficient CF recommendation, which consists of two main phases. In Phase~(1) (graph learning), we first construct a standard, commonly used graph as the base graph (e.g., a $k$NN graph), and then apply a GSP-based method to detect and remove superfluous edges from the original graph. In Phase~(2) (prediction), we perform collaborative filtering on the learned sparse graph to generate recommendations for the active user. In the rest of this section, we provide a detailed description of these two phases.

\subsubsection{Phase 1: Graph Learning}\hfill\break

\textbf{Step (1): Initial graph construction}

Collaborative filtering leverages similarities among users to make recommendations, and these user--user relations can be modeled as a graph. In practice, $k$-nearest-neighbor (kNN) graphs are most commonly used for this purpose. Accordingly, we first construct a kNN graph as an initial representation of user--user relations. A straightforward kNN construction requires $O(|N|^2)$ time, where $|N|$ is the number of users (nodes), which is prohibitive at scale. We therefore employ an approximate $k$NN graph construction algorithm~\cite{malkov2018efficient}, which is based on an extension of a probabilistic skip-list structure and reduces the complexity to $O(|N|\log|N|)$, enabling efficient processing on large datasets.\\

\textbf{Step (2): GSP-based edge pruning}

Probabilistic graphical models (PGMs) are powerful tools for discovering and analyzing structure in complex systems. 
Consider a data matrix $\mathbf{X}\in\mathbb{R}^{n\times m}$, where $n$ is the number of variables (graph nodes) and $m$ is the number of observations (samples). 
From a graph signal processing (GSP) perspective, each column of $\mathbf{X}$ can be viewed as a graph signal defined on the $n$ nodes. 
Under this view, the precision matrix of the corresponding (Gaussian) PGM can be estimated by solving the following optimization problem~\cite{dong2019learning}:
\begin{equation}\label{formula_lasso_dong}
\max_{\Theta \succ 0}\;
F(\Theta)
=\log\det(\Theta)
-\frac{1}{m}\operatorname{Tr}\!\left(\mathbf{X}^{\top}\Theta \mathbf{X}\right)
-\beta\|\Theta\|_{1},
\end{equation}
where $\Theta = L+\frac{1}{\sigma^{2}}I$ is the precision matrix, $L$ denotes the (combinatorial) graph Laplacian, $I$ is the identity matrix, and $\sigma^{2}$ is the prior feature variance. 
$\operatorname{Tr}(\cdot)$ denotes the matrix trace, and $\|\cdot\|_{1}$ is the entry-wise $\ell_{1}$ norm, i.e.,
$\|\Theta\|_{1}=\sum_{i,j}\left|\Theta_{i,j}\right|$.

In this paper, we argue that the optimization problem in~(\ref{formula_lasso_dong}) is well suited to collaborative filtering. 
The first two terms can be interpreted as the log-likelihood of a Gaussian Markov random field (GMRF). 
When $\Theta = L + \frac{1}{\sigma^{2}}I$, the trace term $\operatorname{Tr}\!\left(\mathbf{X}^{\top}\Theta\mathbf{X}\right)$ in~(\ref{formula_lasso_dong}) contains $\operatorname{Tr}\!\left(\mathbf{X}^{\top}L\mathbf{X}\right)$, which measures the smoothness of the signals in $\mathbf{X}$ over the graph~$G$~\cite{kalofolias2016learn}. 
Equivalently,
\begin{equation}
\operatorname{Tr}\!\left(\mathbf{X}^{\top}L\mathbf{X}\right)
=\frac{1}{2}\sum_{i,j}w_{ij}\|\mathbf{x}_i-\mathbf{x}_j\|_2^2,
\end{equation}
where $\mathbf{x}_i$ denotes the $i$-th row of $\mathbf{X}$ (i.e., the interaction/preference vector of user~$i$). 
This formulation encourages users connected by larger edge weights to have more similar interaction (preference) vectors, aligning with the core assumption of collaborative filtering. 
Maximizing~(\ref{formula_lasso_dong}) therefore encourages $\operatorname{Tr}\!\left(\mathbf{X}^{\top}L\mathbf{X}\right)$ to be small, i.e., it promotes signal smoothness with respect to the learned graph. 
The $\beta\|\Theta\|_{1}$ term serves as an $\ell_{1}$ regularizer that enforces sparsity.

However, directly estimating a graph by solving~(\ref{formula_lasso_dong}) is often impractical in large-scale settings, because log-determinant optimization is computationally demanding.
Recently, \cite{wang2022scalable} proposed an efficient indirect approach to approximately solve~(\ref{formula_lasso_dong}).
In this paper, we adopt this indirect strategy to refine the graph topology used in our CF-based recommendation system.
The key idea is to make the dependence of the objective on individual edges explicit, so that we can compute an edge-wise score that quantifies each edge's 11contribution to the objective. 
We then prune edges with the smallest scores to obtain a sparse yet informative graph.

To derive the edge-wise criterion, we start from an edge-based decomposition of the Laplacian. 
As shown in~\cite{mahoney2016lecture}, the graph Laplacian can be expanded as
\begin{equation}\label{wholeEdge}
\mathbf{L}=\sum_{(u,v)\in E} w_{uv}\mathbf{e}_{uv}\mathbf{e}_{uv}^T,
\end{equation}
where $w_{uv}$ denotes the edge weight and $\mathbf{e}_{uv}=\mathbf{e}_u-\mathbf{e}_v$.

\cite{wang2022scalable} shows that, by substituting~(\ref{wholeEdge}) into the precision matrix $\Theta = L + \frac{I}{\sigma^2}$ and rewriting the sparsity term as
${\beta}\|\Theta\|_{1} = \overline{\beta}\sum\limits_{(u,v)\in E} w_{uv} + \alpha$, the optimization problem~(\ref{formula_lasso_dong}) can be reformulated as~(\ref{newObj}).

\begin{equation}\label{newObj}
  \begin{array}{l}
F = \sum\limits_{i = 1}^n\log(\frac{1}{\sigma^2}+\lambda_i)-\frac{1}{n}\frac {Tr(\textbf{X}\textbf{X}^T)}{\sigma^2}-\frac{1}{n}\sum\limits_{(u,v)\in E}w_{uv}\|\textbf{X}^T\mathbf{e}_{uv}\|_2^2\\-\overline{\beta}\sum\limits_{(u,v)\in E}w_{uv}-\alpha.
\end{array}
\end{equation}

where $\lambda_i$ denotes the $i$-th Laplacian eigenvalue, $
{\beta}{{\|\Theta\|}}^{}_{1} = \overline{\beta}\sum\limits_{(u,v)\in E}w_{uv}+\alpha
$ and $\alpha$ denotes a constant number.

Taking partial derivatives with respect to $w_{uv}$ leads to:

\begin{equation}\label{pd3}
\frac{\partial F}{\partial w_{uv}} = (1-\frac{1}{\eta_{uv}})\|U^T\mathbf{e}_{uv}\|_2^2-\overline{\beta},
\end{equation}

where $U$ is a spectral embedding matrix constructed by the nontrivial eigenvalues and eigenvectors of the graph Laplacian: 

\begin{equation}\label{U}
U=
\left[\frac{{u_2}}{\sqrt{\frac{1}{\sigma^2}+\lambda_2}}, ... ,\frac{{u_r}}{\sqrt{\frac{1}{\sigma^2}+\lambda_r}} \right],
\end{equation} and  \begin{equation}\label{eta}
\eta_{uv} = \frac{\|U^T\mathbf{e}_{uv}\|_2^2}{\frac{1}{n}\|\textbf{X}^T\mathbf{e}_{uv}\|_2^2},
\end{equation}

implying that as long as $\eta_{uv}>1$ holds for edge$(u,v)$ and $\overline{\beta}$ is properly selected, the greater $\eta_{uv}$ and $\|U^T\mathbf{e}_{uv}\|_2^2$ values can lead to more significant increase of the objective function.

Starting from an initial base graph, we compute the edge scores implied by~(7) for all existing edges.
We further multiply each edge score by a coefficient, namely the edge weight in the base graph, which we interpret as a confidence measure of the edge: a larger edge weight indicates that the connection between $u$ and $v$ is supported by stronger and more reliable evidence.
With this confidence weighting, the topological importance of low-confidence edges is down-weighted, whereas the topological importance of high-confidence connections is amplified.
Finally, we prune edges according to this confidence-weighted score to obtain a sparse graph.

\subsubsection{Phase 2: Prediction on the Learned Graph}\hfill\break

Given the learned sparse user--user graph $G=(V,E)$ from Phase~1, we perform a standard neighborhood-based collaborative filtering procedure on $G$ to generate recommendations for an active user. 
For a target user $u$ and an item $i$, we predict the preference score by aggregating the feedback of $u$'s neighbors on the learned graph:
\begin{equation}\label{eq:cf_pred}
\hat{r}_{u,i} = \frac{\sum_{v\in \mathcal{N}(u)} w_{uv}\, r_{v,i}}{\sum_{v\in \mathcal{N}(u)} |w_{uv}|},
\end{equation}
where $\mathcal{N}(u)$ denotes the neighbor set of user $u$ in $G$, $w_{uv}$ is the learned edge weight between users $u$ and $v$, and $r_{v,i}$ is user $v$'s observed feedback on item $i$. 
Because the learned graph is substantially sparser than the initial base graph (e.g., a $k$NN graph), the neighborhood aggregation in~(\ref{eq:cf_pred}) becomes significantly more efficient while maintaining competitive recommendation accuracy.

\section{Experiments}

In this section, we conduct extensive experiments to evaluate the performance of the proposed algorithm.

\subsection{Experimental Setup}

\noindent\textbf{Dataset and split.}
We evaluate the proposed method on the MovieLens-100K dataset using the standard split, where the training set contains 80,000 ratings and the test set contains 20,000 ratings. The dataset includes $N=943$ users and $M=1682$ items.\\

\noindent\textbf{Initial graph construction and embedding.}
We construct an initial user--user graph $G_0$ using a union-symmetric $k$NN scheme with $K_{\text{init}}=120$: for each user $u$, we connect $u$ to its $K_{\text{init}}$ most similar users, and then symmetrize the directed $k$NN graph by taking the union of edges, i.e., an undirected edge $(u,v)$ is included if $v\in\mathcal{N}_{K_{\text{init}}}(u)$ or $u\in\mathcal{N}_{K_{\text{init}}}(v)$.
User similarity is computed by mean-centered cosine similarity on the training interaction matrix, and negative similarities are clipped to zero.
Following common practice, we further apply a shrinkage weighting based on co-rated counts:
$w_{uv} = \cos(u,v)\cdot \frac{c_{uv}}{c_{uv}+\texttt{SHRINK}}$ with $\texttt{SHRINK}=50$,
where $c_{uv}$ is the number of co-rated items between users $u$ and $v$.\\
The dimension for spectral embedding is set to $32$.\\

\noindent\textbf{Evaluation metric.}
Mean Absolute Error (MAE) is one of the most commonly used metrics for evaluating rating prediction quality in recommendation systems~\cite{adomavicius2005toward}. 
It measures the average absolute deviation between the predicted ratings and the ground-truth ratings, and is defined as
\begin{equation}\label{eqn:mae}
\mathrm{MAE}=\frac{1}{N}\sum_{i=1}^{N}\left|p_i-q_i\right|,
\end{equation}
where $N$ is the number of rating--prediction pairs, $p_i$ is the rating predicted by the algorithm, and $q_i$ is the corresponding ground-truth rating. 
Lower MAE indicates better prediction accuracy.

\subsection{Experimental Results}

\begin{figure}[!htbp]
\centering\includegraphics[scale=0.7]{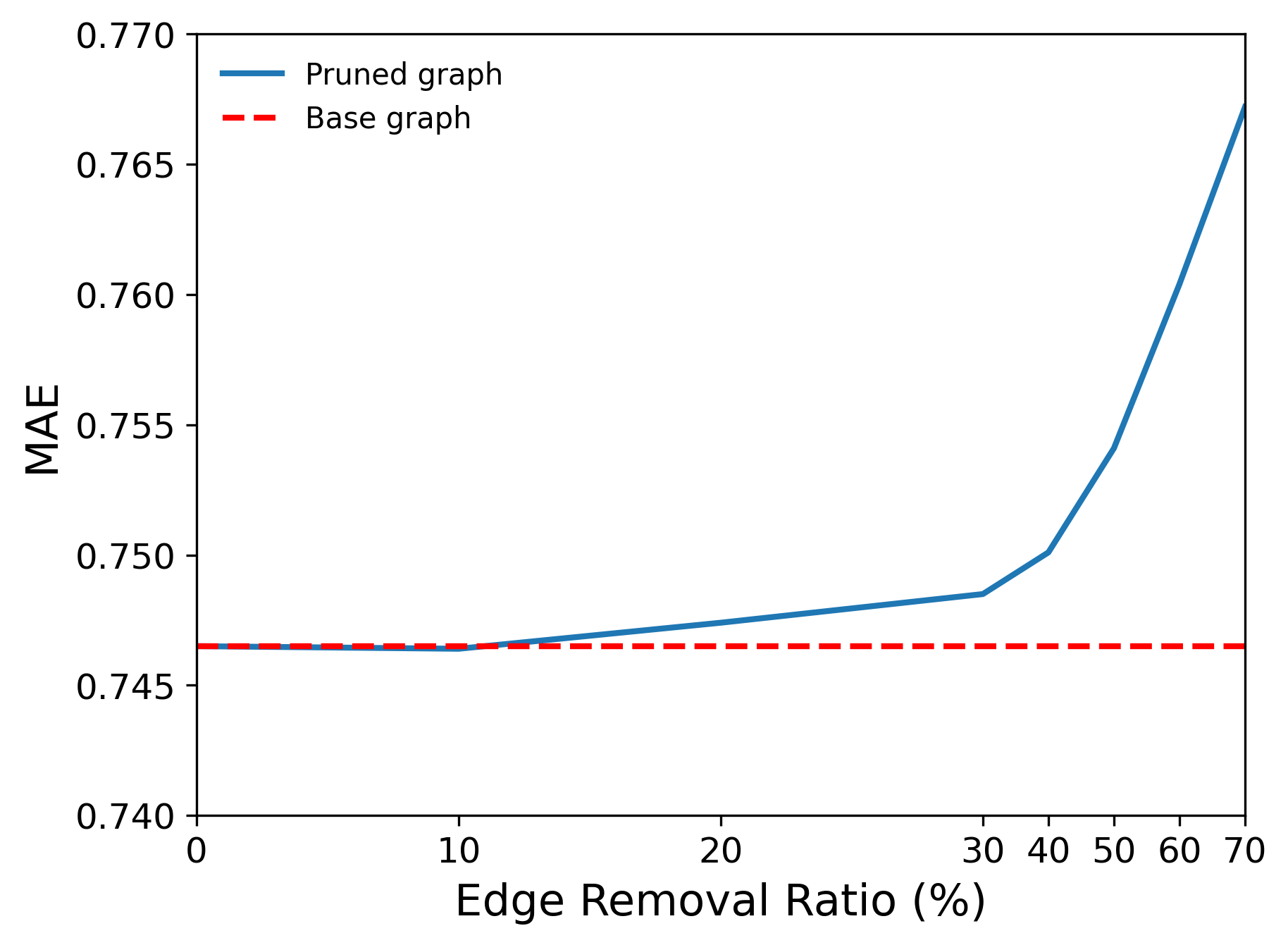}
\caption{MAE vs. Edge Removal Ratio.\protect\label{fig:mae}}
\end{figure}

Table~\ref{table:density} and Figure~\ref{fig:mae} show how the MAE varies as the edge removal ratio increases. We observe that the proposed pruning strategy preserves prediction accuracy well: even after removing 30\% of the edges, the MAE remains essentially the same as that obtained on the original base graph.

\begin{table}[!htbp]
\centering
\setlength{\tabcolsep}{14pt}          
\renewcommand{\arraystretch}{1.15}    
\caption{MAE under Different Edge Removal Ratios}
\label{tab:mae_vs_pruning}
\begin{tabular*}{\linewidth}{@{\extracolsep{\fill}} c c c @{}}
\hline
\textbf{Edge removal ratio} & \textbf{MAE} & \textbf{\#Edges} \\
\hline
0\%  & 0.7465 & 71{,}976 \\
10\% & 0.7464 & 64{,}778 \\
20\% & 0.7474 & 57{,}581 \\
30\% & 0.7485 & 50{,}383 \\
40\% & 0.7501 & 43{,}186 \\
50\% & 0.7541 & 35{,}988 \\
60\% & 0.7604 & 28{,}790 \\
70\% & 0.7672 & 21{,}593 \\
75\% & 0.7727 & 17{,}994 \\
80\% & 0.7803 & 14{,}395 \\
85\% & 0.7898 & 10{,}796 \\
90\% & 0.8074 &  7{,}198 \\
92\% & 0.8150 &  5{,}758 \\
95\% & 0.8301 &  3{,}599 \\
\hline
\end{tabular*}\label{table:density}
\end{table}

\section{Conclusion}

In this paper, we proposed a graph learning framework for efficient collaborative filtering. Starting from a standard user--user $k$NN graph, our method leverages graph signal processing to identify and prune superfluous edges, yielding a substantially sparser topology while maintaining recommendation accuracy. Experiments on benchmark data set demonstrate that a significant portion of edges (e.g., 30\%) can be removed with negligible change in MAE, improving computational efficiency without sacrificing performance.

\end{document}